# MUSCLE FATIGUE DEGRADES FORCE SENSE AT THE ANKLE JOINT

Nicolas VUILLERME and Matthieu BOISGONTIER

Laboratoire TIMC-IMAG, UMR UJF CNRS 5525, Grenoble, France.

* Address for correspondence:

Nicolas VUILLERME

Laboratoire TIMC-IMAG, UMR UJF CNRS 5525,

Faculté de Médecine

38706 La Tronche cédex, France.

Fax: (33) (0) 4 76 51 86 67

Email: nicolas.vuillerme@imag.fr

**Abstract**

To investigate the effects of muscle fatigue on force sense at the ankle joint, ten young healthy adults were asked to perform an isometric contra-lateral force ankle-matching task in two experimental conditions of (1) No-fatigue and (2) Fatigue of the plantar-flexor muscles. Measures of the overall accuracy and the variability of the force matching performances were determined using the absolute error and the variable error, respectively. Results showed less accurate and less consistent force matching performances in the Fatigue than No fatigue condition, as indicated by decreased absolute and variable errors, respectively. The present findings evidence that muscle fatigue degrades force sense at the ankle joint.

**Key-words:** Proprioception; Muscle fatigue; Force matching; Ankle.

**1. Introduction**

In recent years, numerous studies have reported deteriorated postural control during quiet standing following plantar-flexor muscles fatigue (e.g. [1-4]). These observations have been suggested to stem from a decreased ankle proprioception and an inability to produce or sustain required force output with the fatigued plantar-flexor muscles. At this point, although plantar-flexor muscles fatigue has recently been shown to degrade the sense of limb position [5], its effect on the sense of force is yet to be established. The present experiment was thus designed to address this issue.

**2. Methods**

2.1. Subjects

Ten male young healthy adults (age = 25.1 ± 2.9 years) voluntarily participated in the experiment. They gave their informed consent to the experimental procedure as required by the Helsinki declaration (1964) and the local Ethics Committee. None of the subjects presented any history of injury, surgery or pathology to either lower extremity that could affect their ability to perform the experiment.

2.2. Experimental procedure

Subjects were seated comfortably in a chair with their right and left foot put on a pressure mapping system (FSA Seat 32/63, Vista Medical Ltd.), allowing real-time acquisition of the magnitude of pressure and the computation of the force exerted on each left and right foot sole. The ankle and knee joints were locked in place at 10° of plantar-flexion and 110° of flexion respectively. A handheld press-button allowed recording the matching. In addition, a panel was placed above the subject's legs to eliminate visual feedback.

Two different target force levels of (1) 50 N and (2) 150 N were used for all subjects, irrespective of their individual physical capacity, to best simulate real situations (e.g., standing, walking, running, driving). Once the subjects had generated the required target force level (50 N or 150 N) through isometric contractions of their left plantar-flexor muscles, they were asked to match the magnitude of this reference force through isometric contractions of their right plantar-flexor muscles. When they felt that they had reached the target force, they were asked to press the button held in their right hand, thereby registering the matched force. Subjects were not given feedback about their force matching performance and were not given any speed constraints other than 10-s delay to perform one trial. This isometric contra-lateral force ankle-matching task was executed under two states of plantar-flexor muscles fatigue. (1) The No fatigue condition served as a control condition, whereas (2) in the Fatigue condition, the measurements were performed immediately after a fatiguing procedure. Its aim was to induce a muscular fatigue in the plantar-flexor muscles of the right leg until maximal exhaustion [5]. Subjects were asked to perform toe-lifts with their right leg as many times as possible following the beat of a metronome (40 beats/min). Verbal encouragement was given to ensure that subjects worked maximally. The fatigue level was reached when subjects were no more able to complete the exercise. Immediately on the cessation of exercise, the subjective exertion level was assessed through the Borg CR-10 scale [6]. Subjects rated their perceived fatigue in the plantar-flexor muscles as almost "extremely strong" (mean Borg ratings of $7.8 \pm 0.9$ and $8.1 \pm 1.0$, for 50 N and 150 N, respectively). To ensure that measurements in the Fatigue condition were obtained in a real fatigued state, i.e. to limit recovery effect, various rules were respected [5]. (1) The fatiguing exercise took place beside the experimental set-up to minimise the time between the exercise-induced fatiguing activity and the proprioceptive measurements, and (2) the duration of the data collection following the fatiguing exercise lasted approximately 1 minute.

For each force target level of 50 N and 150 N and each experimental condition of No fatigue and Fatigue of the plantar-flexor muscles, subjects performed 5 trials, for a total of 20 trials. The order of presentation of the two target force levels was randomized over subjects.

2.3. Data analysis

Two dependent variables were used to assess force ankle-matching performances [7].

(1) The absolute error (AE), the absolute value of the difference between the force developed by the right matching plantar-flexor muscles and the force developed by the left reference plantar-flexor muscles, is a measure of the overall accuracy of the force ankle-matching performances.

(2) The variable error (VE), the variance around the mean constant error score, is a measure of the variability of the force ankle-matching performances.

Decreased AE and VE scores indicate increased accuracy and consistency of the force ankle-matching performances, respectively [7].

2.4. Statistical analysis

Data obtained for AE and VE were submitted to separate 2 Target force levels (50 N *vs*. 150 N) × 2 Fatigues (No fatigue *vs*. Fatigue) analyses of variances (*ANOVAs*) with repeated measures of both factors. Post hoc analyses (*Newman-Keuls*) were performed whenever necessary. Level of significance was set at 0.05.

3. Results

Figure 1 illustrates the force matching errors from a typical subject measured for each trial executed in the two target force level conditions of 50 N (A) and 150 N (B) and the two conditions of No fatigue (*white squares*) and Fatigue (*black squares*) conditions.

------------------------------------

Please insert Figure 1 about here

------------------------------------

Analysis of the AE showed a main effect of Target force level ($F(1,9)=5.06$, $P<0.05$), yielding higher values in the 150 N than 50 N condition, and a main effect of Fatigue ($F(1,9)=46.39$, $P<0.001$), yielding higher values in the Fatigue than No fatigue condition, (Figure 2A).

------------------------------------

Please insert Figure 2 about here

------------------------------------

Analysis of the VE showed a main effect of Fatigue, yielding higher values in the Fatigue than No fatigue condition ($F(1,9)=12.43$, $P<0.01$, Figure 2B).

## 4. Discussion

Results showed increased AE in the 150 N relative to the 50 N condition, in line with previous studies which also have reported larger force matching errors with larger target force levels [8,9]. More interestingly, larger AE (Figure 2A) and VE (Figure 2B) were observed in the Fatigue than No fatigue condition. In addition to the absence of significant interaction of Force target × Fatigue, these results indicated less accurate and less consistent force matching performances, whatever the target force level, in the Fatigue than No fatigue condition. This result is in accordance with the existing literature reporting degraded force sense following muscle fatigue induced at the elbow [8-12]. Although it is generally accepted that sensation of muscle force arises from two separate sources, (*i*) the sense of tension generated by afferent feedback from the muscle (Golgi tendon organs), and (*ii*) the sense of effort generated centrally (for review see e.g. [13]), it has been proposed the alteration in force sense following muscle fatigue to mainly stem from a change in the sense of effort, that is, an entirely central mechanism (e.g. [12]).

Together with the alteration of ankle joint position sense recently reported following plantar-flexor muscles fatigue [5], the degraded force sense at the ankle joint observed with fatigue in the present experiment could contribute to the adverse effects of plantar-flexor muscles fatigue on postural control reported in the literature (e.g. [1-4]). A study investigating the relationship between ankle proprioceptive acuity and balance ability during quiet standing following exercise-induced plantar-flexor muscles fatigue is planned to address this issue.

**Acknowledgements**

The authors would like to thank subject volunteers. The company Vista Medical is acknowledged for supplying the FSA pressure mapping system. Special thanks also are extended to Olivier Chenu for technical assistance and Zora Boulenger for various contributions.

**References**

1. Ledin T, Fransson PA, Magnusson M. Effects of postural disturbances with fatigued triceps surae muscles or with 20% additional body weight. Gait & Posture 2004;19:184-93.

2. Noda M, Demura S. Comparison of quantitative analysis and fractal analysis of center of pressure based on muscle fatigue. Percept Mot Skills 2006;102:529-542.

3. Vuillerme N, Burdet C, Isableu B, Demetz S. The magnitude of the effect of calf muscles fatigue on postural control during bipedal quiet standing with vision depends on the eye-visual target distance. Gait & Posture 2006;24:169-172.

4. Vuillerme N, Forestier N, Nougier V. Attentional demands and postural sway: the effect of the calf muscles fatigue. Med Sci Sports Exerc 2002;34:1607-1612.

5. Vuillerme N, Boisgontier M, Chenu O, Demongeot J, Payan Y. Tongue-placed tactile biofeedback suppresses the deleterious effects of muscle fatigue on joint position sense at the ankle. Exp Brain Res 2007 183:235-240.

6. Borg G. Psychological scaling with applications in physical work and the perception of exertion. Scand. J. Work Environ. Health 1990;16:55-58.

7. Schmidt RA. Motor Control and Learning, 2nd Edition. Human Kinetics, Champaign, IL; 1988.

8. Proske U, Gregory JE, Morgan DL, Percival P, Weerakkody NS, Canny BJ. Force matching errors following eccentric exercise. Hum Mov Sci 2004;23:365-378.

9. Weerakkody NS, Percival P, Morgan DL, Gregory JE, Proske U. Matching different levels of isometric torque in elbow flexor muscles after eccentric exercise. Exp Brain Res 2003;149:141-150.

10. Brockett C, Warren N, Gregory JE, Morgan DL, Proske U. A comparison of the effects of concentric versus eccentric exercise on force and position sense at the human elbow joint. Brain Res 1997;771:251-258.

11. Carson RG, Riek S, Shahbazpour N. Central and peripheral mediation of human force sensation following eccentric or concentric contractions. J Physiol (Lond) 2002;539:913-925.

12. Jones LA, Hunter IW. Effect of fatigue on force sensation. Exp Neurol 1983;81:640-650.

13. Gandevia SC. Kinaesthesia: roles for afferent signals and motor commands. In: Rowell LB, Shepherd JT, editors. Handbook of physiology, Section 12. Exercise: regulation and integration of multiple systems. New York: Oxford University Press; 1996. p 128-172.

**Figure captions**

**Figure 1.** Illustration of the force matching errors from a typical subject measured for each trial executed in the two target force level conditions of 50 N (A) and 150 N (B) and the two conditions of No fatigue and Fatigue conditions of the plantar-flexor muscles. These two experimental conditions are presented with different symbols: No fatigue (*white squares*) and Fatigue (*black squares*).

**Figure 2.** Mean and standard deviation for the absolute error (A) and variable error (B) for two target force levels of 50 N and 150 N and the two conditions of No fatigue and Fatigue of the plantar-flexor muscles. These two experimental conditions are presented with different symbols: No fatigue (*white bars*) and Fatigue (*black bars*).

**Figure 1**

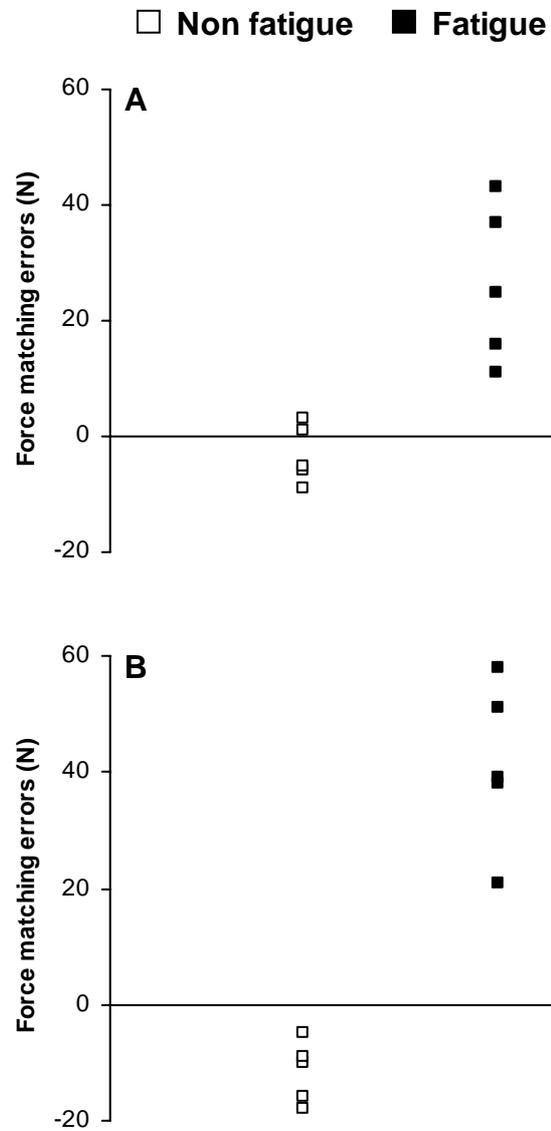

**Figure 2**

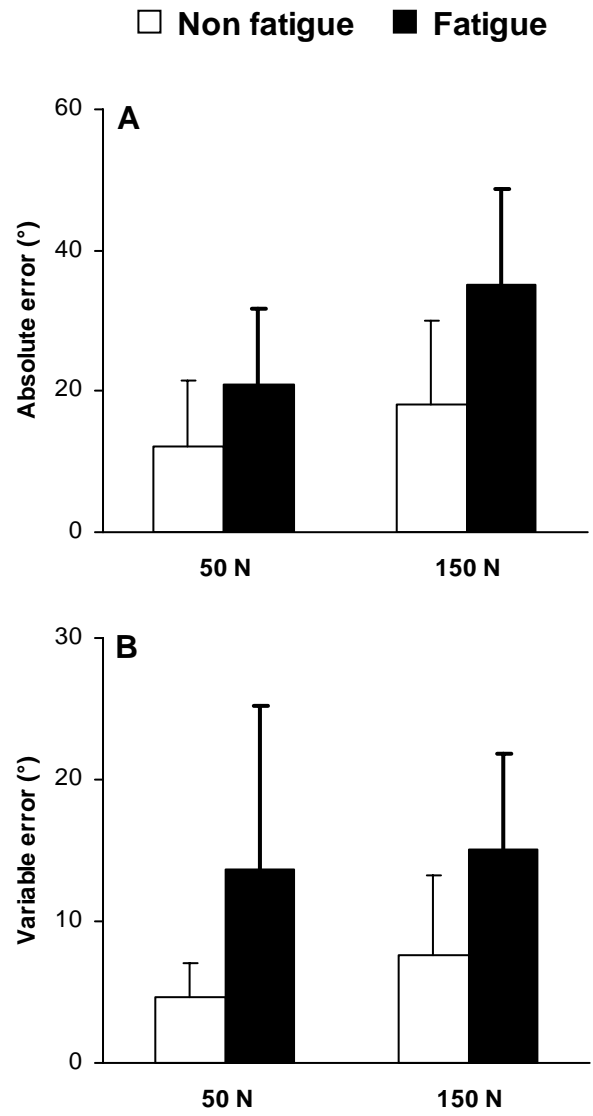